\documentclass{svproc}
\usepackage{url}
\usepackage{mathptmx}       % selects Times Roman as basic font
\usepackage{makeidx}         % allows index generation
\usepackage{graphicx}        % standard LaTeX graphics tool
\usepackage{multicol}        % used for the two-column index
\usepackage[bottom]{footmisc}
\usepackage[cmex10]{amsmath}% places footnotes at page bottom
\usepackage[cmex10]{amsmath}
\usepackage{amssymb}
\usepackage{booktabs}
\usepackage{graphicx}
\usepackage{multirow}
\usepackage{caption}
\usepackage[colorlinks=true,citecolor=black,linkcolor=black,urlcolor=black]{hyperref}
\usepackage{tikz,float,epsf,caption,subcaption}

\begin{document}
% \mainmatter              % start of a contribution
%
\title{Explicit Feedbacks Meet with  Implicit Feedbacks : A Combined Approach for Recommendation System}
\titlerunning{Explicit Feedbacks Meet with Implicit Feedback}  % abbreviated title (for running head)
\author{Supriyo Mandal\inst{1} \and Abyayananda Maiti\inst{2}}
\authorrunning{Supriyo Mandal and Abyayananda Maiti.}
 % \tocauthor{Ivar Ekeland, Roger Temam, Jeffrey Dean, David Grove,
% Craig Chambers, Kim B. Bruce, and Elisa Bertino}
%
\institute{Department of Computer Science and Engineering,\\
Indian Institute of Technology Patna,\\
Patna, Bihar, India 801103,\\
\email{supriyo.pcs17@iitp.ac.in.}
\and
Department of Computer Science and Engineering,\\
Indian Institute of Technology Patna,\\
Patna, Bihar, India 801103,\\
\email{abyaym@iitp.ac.in.}}

\maketitle              % typeset the title of the contribution
\begin{abstract}
Recommender systems recommend items more accurately by analyzing users' potential interest on different brands' items.
%
% By focusing on users' implicit feedbacks and explicit feedbacks, online recommender systems improve the accuracy of users' rating prediction.
%
In conjunction with users' rating similarity, the presence of users' implicit feedbacks like clicking items, viewing items specifications, watching videos etc. have been proved to be helpful for learning users' embedding,
that helps better rating prediction of users.
Most existing recommender systems focus on modeling of ratings and implicit feedbacks  ignoring users' explicit feedbacks.
Explicit feedbacks can be used to validate the reliability of the particular users and can be used to learn about the users' characteristic.
Users' characteristic mean what type of reviewers they are.
In this paper, we explore three different models for recommendation with more accuracy focusing on users' explicit feedbacks and implicit feedbacks.
First one is $RHC-PMF$ that predicts users' rating more accurately based on user's three explicit feedbacks (rating, helpfulness score and centrality)
and second one is $RV-PMF$, where user's implicit feedback (view relationship) is considered.
Last one is $RHCV-PMF$, where both type of feedbacks are considered.
In this model users' explicit feedbacks' similarity indicate the similarity of their reliability and characteristic and implicit feedback's similarity indicates their preference similarity.
Extensive experiments on  real world dataset, $i.e.$ Amazon.com online review dataset  shows that our models perform better compare to base-line models in term of users' rating prediction.
$RHCV-PMF$ model also performs better rating prediction  compare to baseline models for cold start users and cold start items.
% We would like to encourage you to list your keywords within
% the abstract section using the \keywords{...} command.
\end{abstract}

\section{Introduction}

In various domains like E-commerce platforms, online news, online movie sites etc., recommender system performs an important role in attenuating information overburden, having been notoriously adopted. 
Based on  information of demographic profiles and previous preferences of users, recommender systems~\cite{history} predict  users' rating or purchasing decision of items and  recommend right items or right news or suitable friends to the interested users based on prediction.
%
% A high ranked recommendation system can not only capture users' personalized preferences, but also increase both satisfaction for users and revenue for providers.
Most of the existing recommendation approaches could be mainly categorized into Content-Based approach~\cite{CB1} and Collaborative Filtering ($CF$) approach~\cite{CF2}. 
Memory Based method~\cite{MB2}  and Model Based method~\cite{model1}  are two categories of $CF$. 
%
% Memory-based methods mainly analysis the user's neighborhood preference regarding users or items from the user-item rating matrix(utility matrix) while model-based methods generally predict an underlying model that supervises the way users rate items, it demands more satisfactory performance than memory-based model.
%
% In spite of huge popularity of various model-based methods, matrix factorization ($MF$) based model has become one of the most  model-based methods just because of its high predictive accuracy and high scalability~\cite{scale2}. 
% %
% It characterizes both items and users by a shared latent vector space inferred from item rating patterns and users' preferences history~\cite{WMF,wang2015collaborative}.
%
% A lot of research  is going on for elevating $MF$.

To learn high quality user-item embedding, researchers are considering various side informations related to users' implicit feedbacks, user-item interactions~\cite{neural}, product reviews~\cite{conreview,wang2015collaborative} and product images~\cite{park2017also}, that lead to better rating prediction.
%
% Purchasing an item and giving a good or bad rating are two different things.
%
A user purchases an item based on her preference or influenced by others.
After purchasing the item, the user gives positive (good) or negative  (bad) rating based on her satisfaction level.
There are different types of reviewers in online merchandise sites such as positive reviewers, critical reviewers or reliable reviewers.
%
% Rank of positive and  critical reviewers is evaluated based on helpfulness score (discussed in section \ref{help})  
% %
% and top ranking positive and critical reviewers are considered as reliable users who always gives feedback according to the quality of her purchasing item or according to her satisfaction level.
%
As for example, if any user gives bad rating for her purchasing item and more number of users click her review as ``helpful yes", that means she is not only a critical reviewer but also a reliable reviewer because more number of users support her review and rating.
Considering for another case if more number of users click her review as `` helpful no'', that means she is only a critical reviewer but not a reliable one.

In our method, high helpfulness score and centrality score with positive  rating (company should fix threshold value for rating) means the user is not only a positive reviewer but also a reliable reviewer
and low  helpfulness score and low centrality score with positive rating means the user is only a positive reviewer.
Similarly, high helpfulness score and high centrality score with negative  rating means the user is not only a critical reviewer but also a reliable reviewer
and low  helpfulness score and low centrality score with negative  rating means the user is only a negative or critical reviewer.
The company should fix the threshold value to define high, low score range and positive or negative rating range. 
%conreview, park2016trecso,
In \cite{neural,conreview,park2017also,wang2015collaborative} authors are focusing on only implicit feedbacks like user-item interactions, users' view similarity, product images etc. that lead to better user's preference prediction not rating prediction
because a user's rating prediction depends on what type of reviewer she is: positive reviewer or critical reviewer or reliable reviewer.
These type of characteristic, we easily predict from user's explicit feedbacks' similarity.
They ignore users' explicit feedbacks that indicate users' characteristic.

In our research, we consider three explicit feedbacks like ratings, helpfulness score and centrality score and one implicit feedback like view relationship with user-item.
From implicit feedback, we can predict a user's preference areas more accurately and from explicit feedbacks we can guess the user's predicted rating on her preferred items more accurately.
The online merchandise company should recommend such items to a user who not only purchases the item based on her preference but also gives good rating because a user's negative rating or review always affects financial health of a company.

If a user's (positive reviewer) helpfulness score and centrality score is similar to the particular users whose helpfulness score and centrality score are high and rating is positive,
that means the user is a positive and a reliable reviewer and her rating activity will be similar to other positive and reliable reviewers.
Similarly, if a user's (critical reviewer) helpfulness score and centrality score is similar to the particular users whose helpfulness score and centrality score are high and rating is negative,
that means the user is a critical and reliable reviewer and her rating activity will be similar to other critical and reliable reviewers. 
For another case, if a user's helpfulness score and centrality score is similar to the particular users whose helpfulness score and centrality score are low with positive or negative rating, 
that means the user is only positive or critical reviewer, respectively.
So, based on explicit feedbacks similarity, we can predict more accurately a user's characteristic, that helps us to predict her rating activity
and implicit feedback similarity helps us to predict a user's preference areas.
If we consider both explicit and implicit feedback similarity, then we can predict a user's characteristic and preference area both more accurately that help us to predict her ratings activity more precisely.

The rest of the paper is organized as follows: 
In section 2, we describe our methodology that contains, 
Probabilistic Model ($RHC-PMF$) on explicit feedbacks,
Probabilistic Model ($RV-PMF$) on implicit feedback,
and the fusion of explicit feedbacks and implicit feedback.
%
% In this section, we have designed a generalized model for recommendation purpose.
%
In section 3, we show performance of our proposed models and in section 4, we give conclusion and future work.

% \begin{equation}
% \begin{split}
% p(x|\mu, \sigma^{2}) = \mathcal{N}(x \ ;\mu, \sigma^{2}) = \dfrac{1}{\sqrt{2\pi\sigma^{2}}} \ exp \bigg (\dfrac{-(x-\mu)^{2}}{2\sigma^{2}} \bigg )
% \end{split}
% \label{eq:formal}
% \end{equation}

\section {Methodology}

\textbf{Explicit Feedback :}
In this paper, we consider three explicit feedbacks such as user's rating, helpfulness score and centrality score.
% and implicit feedback as view relationship with different items, to predict users' rating more accurately.

\noindent \textbf{ i) Rating Scores :}
In online merchandise site  a user gives rating on her purchased item.
We denote $R_{ij}$ as the rating  of user $U_{i}$ for item $P_{j}$.
For example, Amazon.com ratings span is from $1$ to $5$.
In this paper, we follow the same rating scale.
Rating $1$, $2$ are considered as negating rating and $3$ to $5$ are mentioned as positive rating.
%\subsection{Rating Scores}
% We  scale users' ratings.
%
% We scale ratings from $-2$ to $+2$ i.e. $(-2, -1, 0, +1, +2)$. 
%
%
% So in this case a rating of $5$ would be scaled down to $+2$ and $1$ would be $-2$.
%

\noindent\textbf{ii) Helpfulness Score :} \label{help}
% After purchasing items, user gives ratings and reviews, associated with the items in the online merchandise site. 
Before purchasing an item  users are expected to read the previous reviews regarding the particular item.
In most of the merchandise sites after each review, it asks the question,  ``Was this review helpful to you? (Answer Yes/No)''. 
%
% In this case, the answer to the question corresponds to the feedback on the review. 
%
``Yes'' answer indicates that the review is helpful to the user. 
``No'' answer indicates that the review is not proper or not truthful. 
This helpfulness data can be used to validate the reliability of the particular user's item review. 

We define a new formula to evaluate helpfulness score $H_{ij}$ of the review given by user $U_{i}$ for item $P_{j}$ as follows:
\begin{equation}\label{eq:h}
H_{ij} = (-1)^{\theta}\frac{x_{ij}^{2}}{y_{ij}},
\end{equation} 
where $x_{ij}$ is the number of users who marked the review given by user $U_{i}$ for item $P_{j}$ as helpful and $y_{ij}$ is the total number of users who have answered that question (total count of yes and no).
%
% Our rating scale in the range of [+1, +5].
%
If $R_{ij}$ is in the range of [ +3, +5 ], $\theta$ is equal to 2, otherwise 1. 
The above equation is quadratic in nature because we want to give more priority to the particular users who have more number of ``yes click''.
Helpfulness score with positive sign indicates that the user is a positive reviewer and negative sign indicates that the user is a critical reviewer.
It is very difficult to get the exact information about the users who not only marked  the review as helpful but also purchase the item.
So we do not consider this scenario.
We assume that the users who marked the review as helpful are interested to purchase the item.
In the real world, there are many users who read the reviews of previous users but do not click on helpful ``yes" or ``no".
Our experimental dataset does not provide such type of information.
So, it is out of our consideration.

\noindent\textbf{iii) Review Network and Centrality Score :}
\begin{figure}[hbtp]
\centering
\includegraphics[scale=.5]{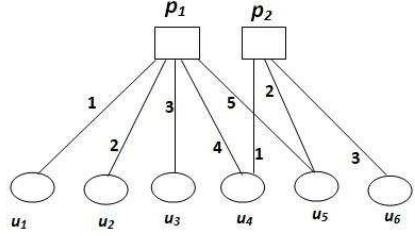}
\caption{Bipartite network between items and users. An edge denotes a review is written by a user on an item.}
\label{fig:review}
\end{figure}
% Since most of the social network data are private, it is hard to evaluate the influence, a user has in her social network related to certain online purchase.
% %
% If that user writes a review on that purchase item in the merchandise site then the influence of that purchase can be measured as the review data are publicly available. 
% %
When we are going to purchase any item from online merchandise sites, we read previous customers' reviews which are related to that particular item.
We can build a network of reviewers based on their purchasing items and timestamps.
We name this network as \textit{Review Network}\cite{mandal2018social}. 

Fig.~\ref{fig:review} depicts a bipartite network presenting the review data set.
Two sets of nodes in that bipartite network are item set and user set.
If user $U_i$ writes a review on item $P_j$ then there will be an edge between them. 
Essentially each edge represents a distinct review.
Notice that, each review has a time stamp of its creation. 
We identify each edge by a unique number that indicates logical time stamp of edge creation. 
Please note that in this figure we have not specified original time stamps. 
For depiction purpose, we have assumed some time direction in Fig.~\ref{fig:review}. 
The edges between user $U_1$ - item $P_1$ and user $U_2$ - item $P_1$ are identified by time stamp $1$ and $2$ respectively, that means user $U_2$ posts her rating  after user $U_1$.
According to the review post timing, here we assume that user $U_2$ purchases the item  after user $U_1$.
In Fig.~\ref{fig:review}, $P_1$, $P_2$ are items and $U_1$, $U_2$, $U_3$, $U_4$, $U_5$, $U_6$  are users who have written reviews on one or two items.    

% Fig.~\ref{fig:review} depicts a bipartite network presenting the review data set.
% %
% Two sets of nodes in that bipartite network are item set and user set.
% %
% If user $u_i$ writes a review on item $p_j$ then there will be an edge between them. 
% %
% Essentially each edge represents a distinct review.
% %
% Notice that, each review has a time stamp of its creation. 
% %
% We identify each edge by unique number that indicates logical time stamp of edge creation. 
% %
% Please note that here we have not specified original time stamps. 
% %
% For depiction purpose we have assumed some time direction in Fig.~\ref{fig:review}. 
%
% The edges between user $u_1$ item $p_1$ and user $u_2$  item $p_1$ are identified by time stamps $1$ and $2$ respectively, that means user $u_2$ purchases item $p_1$ after user $u_1$. 
% %
% In Fig.~\ref{fig:review}, $p_1$, $p_2$ are items of a brand and $u_1$, $u_2$, $u_3$, $u_4$, $u_5$, $u_6$  are users have written reviews on one or more items from those two items.

% We use centrality score of a user from review network to evaluate her influence of promoting certain item.
%
% There are many different kinds of centrality measures such as degree centrality, eigenvector centrality, katz centrality etc. %
%
% It depends on the objective of the company's management which centrality measure they would consider.
%
In this paper, we use degree centrality to evaluate centrality score of each user for a particular item in review network.
We evaluate a user's centrality score based on how many other users read her review.
Our dataset does not provide exact information about the users who read others reviews regarding a particular item.
When a user wants to read the previous reviews of other users for a particular item, in online merchandise site there are two option : a) Most recent reviews b) Top ranking reviews.
Based on our realistic assumption, centrality score of each user is evaluated.
% Top ranking reviews are selected based on helpfulness score.

Most recent reviews are selected based on current time stamp and the time stamp  when the users post reviews. 
Based on most recent reviews centrality score ($most_{ij}$) of user $U_{i}$ for item $P_{j}$ is evaluated from our proposed  equation:  
%
% \begin{equation}\label{eq:m}
% most _{ij}= \sum_{r=1}^{n-i}\dfrac{1}{r} ,
% \end{equation}

%\noindent\begin{minipage}{.5\linewidth}
\begin{equation}\label{eq:m}
most _{ij}= \sum_{r=1}^{n-i}\dfrac{1}{r} ,
\end{equation}
%\end{minipage}%
% \begin{minipage}{.5\linewidth}
% \begin{equation}\label{eq:t}
% top_{ij} = (\dfrac{1}{rk_{i}})^{2}*(n-i),
% \end{equation}
% \end{minipage}
% \begin{minipage}{.34\linewidth}
% \begin{equation}\label{eq:t}
% H_{ij} = (-1)^{\theta}\frac{x_{ij}^{2}}{y_{ij}},
% \end{equation}
% \end{minipage}

where $n$ is total number of users who purchase item $P_{j}$.
$i$ starts from 1 means $U_{1}$ is the first user who purchases  item $P_{j}$.
As for example, in Fig.~\ref{fig:review}  $U_{1}$ is the first user who writes review regarding item $P_{1}$.
Before purchasing the same item, $U_{2}$ reads $U_{1}$'s review as most recent review and for $U_{2}$, $U_{1}$'s centrality score = 1.
When $U_{3}$ purchases the same item, she may read $U_{1}$'s review as second most recent review and for $U_{3}$, $U_{1}$'s  centrality score = 1/2.
Total centrality score of $U_{1}$ user for item $P_{1}$ = $\sum_{r=1}^{4}\dfrac{1}{r}$.

Now we consider centrality score of users based on top ranking reviews. 
We evaluate top ranking reviews  based on their helpfulness score  for  all positive reviewers and all critical reviewers separately from our proposed equation :
\begin{equation}\label{eq:t}
top_{ij} = (\dfrac{1}{k_{i}})^{2}*(n-i),
\end{equation}
where,  $top_{ij}$ is the centrality score of user $U_{i}$ for item $P_{j}$ based on top ranking review.
$k_{i}$ is the ranking for user $U_{i}$.
Higher helpfulness score means rank is higher.
$(n-i)$ indicates how many users purchase the same item after $i^{th}$ user.

Total centrality score $D_{ij}$ is calculated based on our proposed technique as follows:
\begin{equation}\label{eq:c}
%\begin{split}
D_{ij}= (-1)^  {\theta}\times (\alpha*top_{ij}+(1-\alpha)*most_{ij}),
%\end{split}
\end{equation}

where $\alpha$ is weightage value.
The company's management would decide the exact weightage value.
Here we set weightage = 0.5.
% Our rating scale is in the range of [-, +2].
%
If $R_{ij}$ is in the range of [ +3, +5 ], $\theta$ is equals to 2, otherwise 1. 
Centrality score with negative sign indicates that the user is critical reviewer and positive sign for positive reviewer.

\noindent\textbf{Implicit Feedback :}
In this paper, we consider view relationship as implicit feedback between user-item to predict a user's preference areas more accurately.

\noindent\textbf{ i) View Relationship :}
Before purchasing any item, users always view different items on the same category based on their preference areas.
The online merchandise sites have information about the view history of registered users, that helps us to understand the preference area of a user.
Based on this information, recommender system recommends items to users based on their interest.

% We use the matrix $V$= [$V_{ij}$]$ _{n \times m }$ $\in$ $\mathbb{R}^{n \times m}$ to indicate the user-item view matrix produced by the users who view different  items, where $V_{ij}$ is the view score that is given by user $U_{i}$ on item $P_{j}$.
% %
% $V_{ij}$ is equals to 1, if user $U_{i}$ views item $P_{j}$ based on her interest otherwise, 0

% \section{Modeling on Explicit Feedback}

\noindent\textbf{Probabilistic Model with Explicit Feedbacks (RHC-PMF) : } \label{RHC-PMF}

In this section, we first describe how ratings $R$ is approximated by helpfulness score $H$ and centrality scores $D$.
In this section, we introduce user-item rating matrix. 
Typically there are three type of objects, namely users, ratings and items. 
Suppose, $U$ = $\lbrace$ $U_{1}$, $U_{2}$, ...., $U_{n}$ $\rbrace$ be the set of users, ${P}$ = $\lbrace$ $P_{1}$, $P_{2}$, ...., $P_{m}$ $\rbrace$ be the set of items, ${R}$ = $\lbrace$ $R_{1}$, $R_{2}$, ...., $R_{N}$ $\rbrace$ be the set of ratings,
${H}$ = $\lbrace$ $H_{1}$, $H_{2}$, ...., $H_{N}$ $\rbrace$ be the set of helpfulness scores
and ${D}$ = $\lbrace$ $D_{1}$, $D_{2}$, ...., $D_{N}$ $\rbrace$ be the set of centrality scores
where $n,m$ are the number of users and items respectively.
$N$ is the number of rating, helpfulness score and centrality individually.

We use the matrix $R$ = [$R_{ij}$]$_{n \times m }$ $\in$ $\mathbb{R}^{n \times m}$ to indicate the user-item rating matrix produced by the users who give ratings on different purchasing items, where $R_{ij}$ is the rating score that is given by user $U_{i}$ on item $P_{j}$. 
Similarly, the matrix $H$ = [$H_{ij}$]$ _{n \times m }$ $\in$ $\mathbb{R}^{n \times m}$ to indicate the user-item helpfulness score based matrix gained by the users  on different purchasing items, where $H_{ij}$ is the helpfulness  score that is scored by user $U_{i}$ on item $P_{j}$.
Another matrix  $D$ = [$D_{ij}$]$ _{n \times m }$ $\in$ $\mathbb{R}^{n \times m}$ to indicate the user-item centrality score based matrix gained by the users  on different purchasing items, where $D_{ij}$ is centrality  score that is scored by user $U_{i}$ on item $P_{j}$.

% We map the rating $r_{ij}$ (range -2 to +2), $h_{ij}$ (range is varied from brand to brand) and $d_{ij}$ (range is also varied) into an interval (-1,+1) using Eq. \ref{scale}.
%
% \begin{equation}
% x'=c+(x-a)\bigg(\dfrac{d-c}{b-a}\bigg),
% \label{scale}
% \end{equation}
% where x is a value into an interval [a,b] and we have to map it into an interval [c,d].
% 
% $x^{'}$ is scaled value of x into the interval [c,d].
%
% We map it to provide flexibility in our models specially $MLP@SPS$ model.

Our method try to factorize the  rating matrix  $R$ $\in$ $\mathbb{R}^{n \times m}$ into two matrices $W$ $\in$ $\mathbb{R}^{K \times n}$ and $Z$ $\in$ $\mathbb{R}^{K \times m}$.
$W$ is the user latent factor matrix with each column $W_{i}$ being $W_{i}$'s latent feature vector  that indicates how user $U_{i}$'s taste is similar to other users based on ratings and $Z$ is the item latent factor matrix with each column $Z_{j}$ being the latent feature vector  of the item $P_{j}$ that indicates how the other users rate on item $P_{j}$. 
The  helpfulness score based matrix  $H$ $\in$ $\mathbb{R}^{n \times m}$ is also factorized  into two matrices $E$ $\in$ $\mathbb{R}^{K \times n}$ and $F$ $\in$ $\mathbb{R}^{K \times m}$
where, $E$ is the user latent factor matrix with each column $E_{i}$ being $E_{i}$'s latent feature vector  that indicates how user $U_{i}$'s helpfulness score is similar to other users and $F$ is the item latent factor matrix with each column $F_{j}$ being the latent feature vector  of the item $P_{j}$ that indicates how the other users get helpfulness score on item $P_{j}$.
Similarly, the  centrality score based matrix  $D$ $\in$ $\mathbb{R}^{n \times m}$ is factorized into two matrices $C$ $\in$ $\mathbb{R}^{K \times n}$ and $O$ $\in$ $\mathbb{R}^{K \times m}$.
Here we consider users' explicit feedbacks like ratings, helpfulness score and centrality and try to represent each user and item by a low-dimensional vector. 
MF tries to map both users and items to a joint latent factor space with low-dimensionally $K$ such that user-item interactions are modeled as inner items in that space.

We define the conditional distribution~ \cite{mnih2008probabilistic} over the observed ratings as:
\begin{equation}
\begin{split}
p(R|W, Z, \sigma_{R}^{2}) = \prod_{i=1}^{n}\prod_{j=1}^{m}
\bigg [\mathcal{N} \bigg (R_{ij}|g(W_{i}^{T}Z_{j}) \bigg ),  \sigma_{R}^{2} \bigg]^{I_{ij}^{R}},  
\end{split}
\label{eq:R1}
\end{equation}

where $\mathcal{N}(x| \mu, \sigma^{2})$ denotes the probability function of a Gaussian distribution with mean $\mu$ and variance $\sigma^{2}$.
$I_{ij}^{R}$ is equal to 1 if user $U_{i}$ rated item $P_{j}$, and 0 otherwise.
Tanh function $g(.)$ is used to map the range of $(W_{i}^{T}Z_{j})$ within [-1,+1], and we map  $R_{ij}$ into the same range [-1,+1] using the following Eq. \ref{scale}.
\begin{equation}
x'=c+(x-a)\bigg(\dfrac{d-c}{b-a}\bigg),
\label{scale}
\end{equation}
where x is a value into an interval [a,b] and we have to map it into an interval [c,d].
$x^{'}$ is scaled value of x into the interval [c,d].

For each hidden variable, we place zero-mean spherical  Gaussian priors~\cite{dueck2004probabilistic} as follows :
\begin{equation}
\begin {split}
p(W|\sigma_{W}^{2}) = \prod_{i=1}^{n}
\mathcal{N} \bigg(W_{i}| \ 0, \dfrac{\sigma_{W}^{2}}{n_{w_{i}}}\textbf{I} \bigg ) \ and \ p(Z|\sigma_{Z}^{2}) = \prod_{j=1}^{m}
\mathcal{N} \bigg(Z_{j}| \ 0, \dfrac{\sigma_{Z}^{2}}{n_{z_{j}}}\textbf{I} \bigg ) 
\end{split}
\label{eq:R2}
\end{equation}
Please note that, we do not consider a uniform variance of all users as shown in Eq.\ref{eq:R2}.
We try to make more reasonable to characterize different users with different prior variance for better recommendation.
Here $n_{w_{i}}$ is the number of ratings given by user $U_{i}$ means we have to adjust user $U_{i}$'s prior variance according to $n_{w_{i}}$.
The more number of ratings given by user will contribute more accuracy to learn her rating activity and consequently, the smaller the uncertainty.
This means that the prior variance of user $U_{i}$ will be inversely proportional to  $n_{w_{i}}$.
$n_{z_{j}}$ is the number of users who rate on item $P_{j}$.
%
% \begin{equation}
% \begin{split}
% p(Z|\sigma_{Z}^{2}) = \prod_{j=1}^{m}
% %
% \mathcal{N} \bigg(Z_{j}| \ 0, \dfrac{\sigma_{Z}^{2}}{n_{z_{j}}}\textbf{I} \bigg ) 
% \end{split}
% \label{eq:rating}
% \end{equation}

% \subsection{Probabilistic Model with Helpfulness Score }

Similar as Eq.~\ref{eq:R1},~\ref{eq:R2}  we have defined the another conditional distribution over the observed helpfulness score, that is  $p(H|E, F, \sigma_{H}^{2})$,
where $p(E|\sigma_{E}^{2})$ and $p(F|\sigma_{F}^{2})$,  the priors of the user and item features based on helpfulness score, are modeled as zero-mean spherical Gaussian distribution.  
Due to space limitation, expressions are omitted.
Here $H_{ij}$ is also mapped into the range [-1, +1] using Eq.~\ref{scale}.
% \begin{equation}
% \begin{split}
% p(H|E, F, \sigma_{H}^{2}) 
% = \prod_{i=1}^{n}\prod_{j=1}^{m}
% %
% \bigg [\mathcal{N} \bigg (H_{ij}|g(E_{i}^{T}F_{j}) \bigg ), \sigma_{H}^{2} \bigg]^{I_{ij}^{H}},
% \end{split}
% \label{eq:H1}
% \end{equation}
% %
% \begin{equation}
% \begin{split}
% p(E|\sigma_{E}^{2}),  
% = \prod_{i=1}^{n}
% %
% \mathcal{N} \bigg(E_{i}| \ 0, \dfrac{\sigma_{E}^{2}}{n_{e_{i}}}\textbf{I} \bigg ) , 
% p(F|\sigma_{F}^{2}) 
% % = \prod_{j=1}^{m}
% % %
% % \mathcal{N} \bigg(F_{j}| \ 0, \dfrac{\sigma_{F}^{2}}{n_{f_{j}}}\textbf{I} \bigg ) ,
% \end{split}
% \label{eq:H2}
% \end{equation}
%
Similarly, here we do not consider a uniform variance.
%
% $n_{e_{i}}$ is the total number of ratings for which user $U_{i}$'s helpfulness score does not equal to zero and $n_{f_{j}}$ is the total number of users who gain helpfulness score greater than zero for the rating of item $P_{j}$.
%
% \begin{equation}
% \begin{split}
% p(F|\sigma_{F}^{2}) = \prod_{j=1}^{m}
% %
% \mathcal{N} \bigg(F_{j}| \ 0, \dfrac{\sigma_{F}^{2}}{n_{f_{j}}}\textbf{I} \bigg ) 
% \end{split}
% \label{eq:rating}
% \end{equation}
% \subsection{Probabilistic Model with Centrality Score}
%
Another conditional distribution over the observed centrality score is
$p(D|C, O, \sigma_{D}^{2})$ where $p(C|\sigma_{C}^{2})$ and $p(O|\sigma_{O}^{2})$, the priors of the user and item features based on centrality score, are modeled as zero-mean spherical Gaussian distribution.
% \begin{equation}
% \begin{split}
% p(D|C, O, \sigma_{D}^{2}) = \prod_{i=1}^{n}\prod_{j=1}^{m}
% %
% \bigg [\mathcal{N} \bigg (D_{ij}|g(C_{i}^{T}O_{j}) \bigg ), \sigma_{D}^{2} \bigg]^{I_{ij}^{D}}
% \end{split}
% \label{eq:C1}
% \end{equation}
% %
% %
% For each hidden variable, we place zero-mean spherical  Gaussian priors as follows :
% \begin{equation}
% \begin{split}
% p(C|\sigma_{C}^{2}) = \prod_{i=1}^{n}
% %
% \mathcal{N} \bigg(C_{i}| \ 0, \dfrac{\sigma_{C}^{2}}{n_{c_{i}}}\textbf{I} \bigg ) , p(O|\sigma_{O}^{2}) = \prod_{j=1}^{m}
% %
% \mathcal{N} \bigg(O_{j}| \ 0, \dfrac{\sigma_{O}^{2}}{n_{o_{j}}}\textbf{I} \bigg ). 
% \end{split}
% \label{eq:C2}
% \end{equation}
%
% Here $n_{c_{i}}$ is the total number of ratings for which user $U_{i}$'s centrality score does not equal to zero and $n_{o_{j}}$ is the total number of users who gain centrality score does not equal to zero for the rating of item $P_{j}$.

In our model user vector $W_{i}$ based on rating approximates to the user vector $E_{i}$ based on helpfulness score.
So, we define a conditional distribution of $W_{i}$ given $E_{i}$ as follows:
\begin{equation}
\begin{split}
p(W|E,\sigma_{WE}^{2}) = \prod_{i=1}^{n}
\mathcal{N} \bigg(W_{i}| \ E_{i}, {\sigma_{WE}^{2}} \textbf{I} \bigg ) ,
\end{split}
\label{eq:WE}
\end{equation}

where variance $\sigma_{WE}^{2}$ controls the extent by which $W_{i}$ approximates $E_{i}$.
Similarly, we define  another conditional distribution of $W_{i}$ given $C_{i}$, that is $p(W|C,\sigma_{WC}^{2})$,
% \begin{equation}
% \begin{split}
% p(W|C,\sigma_{WC}^{2}) = \prod_{i=1}^{n}
% %
% \mathcal{N} \bigg(W_{i}| \ C_{i}, {\sigma_{WC}^{2}}\textbf{I} \bigg ) ,
% \end{split}
% \label{eq:WC}
% \end{equation}
%
where variance $\sigma_{WC}^{2}$ controls the extent by which user vector $W_{i}$ based on rating approximates $C_{i}$ based on centrality score.
% \begin{equation}
% \begin{split}
% p(O|\sigma_{O}^{2}) = \prod_{j=1}^{m}
% %
% \mathcal{N} \bigg(O_{j}| \ 0, \dfrac{\sigma_{O}^{2}}{n_{o_{j}}}\textbf{I} \bigg ) 
% \end{split}
% \label{eq:rating}
% \end{equation}
%

Now, we can compute the posterior distribution over the hidden variables considering three explicit feedbacks:
\begin{equation}
\begin{split}
p (W, Z, E, F, C, O| R, H, D, \sigma_{R}^{2}, \sigma_{H}^{2}, \sigma_{D}^{2},\sigma_{WE}^{2},\sigma_{WC}^{2}, \sigma_{W}^{2}, \sigma_{Z}^{2}, \sigma_{E}^{2}, \sigma_{F}^{2}, \sigma_{C}^{2}, \sigma_{O}^{2}) 
\propto\\
p(R| W, Z, \sigma_{R}^{2}) %
p(H| E, F, \sigma_{H}^{2}) %
p(D| C, O, \sigma_{D}^{2}) %
p(W| E, \sigma_{WE}^{2}) % 
p(W| C, \sigma_{WC}^{2}) \\
p(W| \sigma_{W}^{2}) %
p(Z| \sigma_{Z}^{2}) %
p(E| \sigma_{E}^{2}) %
p(F| \sigma_{F}^{2}) %
p(C| \sigma_{C}^{2}) %
p(O| \sigma_{O}^{2}). \\
 %
%  =\prod_{i=1}^{n}\prod_{j=1}^{m}
% %
% \bigg [\mathcal{N} \bigg (R_{ij}|g(W_{i}^{T}Z_{j}) \bigg ),  \sigma_{R}^{2} \bigg]^{I_{ij}^{R}} \times \prod_{i=1}^{n}\prod_{j=1}^{m}
% %
% \bigg [\mathcal{N} \bigg (H_{ij}|g(E_{i}^{T}F_{j}) \bigg ), \sigma_{H}^{2} \bigg]^{I_{ij}^{H}}\\
% %
% \times  \prod_{i=1}^{n}\prod_{j=1}^{m}
% %
% \bigg [\mathcal{N} \bigg (D_{ij}|g(C_{i}^{T}O_{j}) \bigg ), \sigma_{D}^{2} \bigg]^{I_{ij}^{D}}
% %
% \times
% \prod_{i=1}^{n}
% %
% \mathcal{N} \bigg(W_{i}| \ E_{i}, {\sigma_{WE}^{2}}\textbf{I} \bigg )
% %
% \times
% \prod_{i=1}^{n}
% %
% \mathcal{N} \bigg(W_{i}| \ C_{i}, {\sigma_{WC}^{2}}\textbf{I} \bigg )\\
% %
% \times
% \prod_{i=1}^{n}
% %
% \mathcal{N} \bigg(W_{i}| \ 0, \dfrac{\sigma_{W}^{2}}{n_{w_{i}}}\textbf{I} \bigg)
% %
% \times
% %
% \prod_{j=1}^{m}
% %
% \mathcal{N} \bigg(Z_{j}| \ 0, \dfrac{\sigma_{Z}^{2}}{n_{z_{j}}}\textbf{I} \bigg)
% %
% \times
% %
% \prod_{i=1}^{n}
% %
% \mathcal{N} \bigg(E_{i}| \ 0, \dfrac{\sigma_{E}^{2}}{n_{e_{i}}}\textbf{I} \bigg)
% %
% \times
% %
% \prod_{j=1}^{m}
% %
% \mathcal{N} \bigg(F_{j}| \ 0, \dfrac{\sigma_{F}^{2}}{n_{f_{j}}}\textbf{I} \bigg)\\
% %
% \times
% %
% \prod_{i=1}^{n}
% %
% \mathcal{N} \bigg(C_{i}| \ 0, \dfrac{\sigma_{C}^{2}}{n_{c_{i}}}\textbf{I} \bigg)
% %
% \times
% %
% \prod_{j=1}^{m}
% %
% \mathcal{N} \bigg(O_{j}| \ 0, \dfrac{\sigma_{O}^{2}}{n_{o_{j}}}\textbf{I} \bigg)
 \end{split}
\label{eq:RVD}
\end{equation}
%
% \section {Modeling on Implicit Feedback}

\noindent\textbf{Probabilistic Model with Implicit Feedback (RV-PMF) :} \label{RV-PMF}

In this section, we first describe how ratings $R$ is approximated by view relationship $V$ between users and items.
In this section, we introduce view relationship between user-item. 
Typically there are three type of objects, namely users, items and view-score. 
Suppose, $U$ = $\lbrace$ $U_{1}$, $U_{2}$, ...., $U_{n}$ $\rbrace$ be the set of users, ${P}$ = $\lbrace$ $P_{1}$, $P_{2}$, ...., $P_{m}$ $\rbrace$ be the set of items, ${V}$ = $\lbrace$ $V_{1}$, $V_{2}$, ...., $V_{N}$ $\rbrace$ be the set of view-score.
We use the matrix $V$= [$V_{ij}$]$ _{n \times m }$ $\in$ $\mathbb{R}^{n \times m}$ to indicate the user-item view matrix produced by the users who view different  items, where 
view score $V_{ij}$ is equals to 1, if user $U_{i}$ views item $P_{j}$ based on her interest otherwise 0.

Our method try to factorize the  view matrix  $V$ $\in$ $\mathbb{R}^{n \times m}$ into two matrices $S$ $\in$ $\mathbb{R}^{K \times n}$ and $U$ $\in$ $\mathbb{R}^{K \times m}$.
In this section, we have defined the  conditional distribution over the observed view items of different users
that is $p(V|S, U, \sigma_{V}^{2})$, where $p(S|\sigma_{S}^{2})$ and $p(U|\sigma_{U}^{2})$, the priors of the user and item features, are modeled as zero-mean spherical Gaussian distribution. 
 Here, $V_{ij}$ is also mapped -1 to +1 scale.
% \begin{equation}
% \begin{split}
% p(V|S, U, \sigma_{V}^{2}) = \prod_{i=1}^{n}\prod_{j=1}^{m}
% %
% \bigg [\mathcal{N} \bigg (V_{ij}|g(S_{i}^{T}U_{j}) \bigg ), \sigma_{V}^{2} \bigg]^{I_{ij}^{V}}
% \end{split}
% \label{eq:V1}
% \end{equation}
% %
% %
% %
% \begin{equation}
% \begin{split}
% p(S|\sigma_{S}^{2}) = \prod_{i=1}^{n}
% %
% \mathcal{N} \bigg(S_{i}| \ 0, \dfrac{\sigma_{S}^{2}}{n_{s_{i}}}\textbf{I} \bigg ), p(U|\sigma_{U}^{2}) = \prod_{j=1}^{m}
% \mathcal{N} \bigg(U_{j}| \ 0, \dfrac{\sigma_{U}^{2}}{n_{u_{j}}}\textbf{I} \bigg )  
% \end{split}
% \label{eq:V2}
%\end{equation}
%

In our model user vector $W_{i}$ based on rating approximates to the user vector $S_{i}$ based on  her view activity of different items.
So, we define a conditional distribution of $W_{i}$ given $S_{i}$, that is $p(W|S,\sigma_{WS}^{2})$
%
% \begin{equation}
% \begin{split}
% p(W|S,\sigma_{WS}^{2}) = \prod_{i=1}^{n}
% %
% \mathcal{N} \bigg(W_{i}| \ S_{i}, {\sigma_{WS}^{2}}\textbf{I} \bigg ) ,
% \end{split}
% \label{eq:WS}
% \end{equation}
%
where variance $\sigma_{WS}^{2}$ controls the extent by which $W_{i}$ approximates $S_{i}$.
%
%
% \begin{equation}
% \begin{split}
% p(U|\sigma_{U}^{2}) = \prod_{j=1}^{m}
% %
% \mathcal{N} \bigg(U_{j}| \ 0, \dfrac{\sigma_{U}^{2}}{n_{u_{j}}}\textbf{I} \bigg ) 
% \end{split}
% \label{eq:rating}
% \end{equation}
We can compute the posterior distribution over the hidden variables based on implicit feedback:
\begin{equation}
\begin{split}
p (W, Z, S, U| R, V, \sigma_{R}^{2}, \sigma_{V}^{2},\sigma_{WS}^{2}, \sigma_{W}^{2}, \sigma_{Z}^{2}, \sigma_{S}^{2}, \sigma_{U}^{2}) 
\propto\\
p(R| W, Z, \sigma_{R}^{2}) %
p(V| S, U, \sigma_{V}^{2}) 
% \\
%
p(W| S, \sigma_{WS}^{2}) %
p(W| \sigma_{W}^{2}) %
p(Z| \sigma_{Z}^{2}) %
p(S| \sigma_{S}^{2}) %
p(U| \sigma_{U}^{2}). %
\end{split}
\label{eq:rating}
\end{equation}
\textbf{RHCV-PMF Model---Fusion of Explicit Feedback and Implicit Feedback :}

% In section \ref{RHC-PMF}, we have designed a model named as $RHC-PMF$ and , where user's rating are associated with her helpfulness score and centrality score.
% %
% We have described another model named as $RV-PMF$ in section \ref{RV-PMF}, where user's ratings are associated with her view activity for various items based on her preference.
%
In this section, we fuse both model ($RHC-PMF$ and $RV-PMF$) and design a model named as $RHCV-PMF$ where user's rating are associated with her explicit feedback (helpfulness score and centrality) and implicit feedback (view activity).

The posterior distribution over the features of users and item based on explicit feedbacks and implicit feedback is given by:
\begin{equation}
\begin{split}
p (W, Z, E, F, C, O, S, U| R, H, D, V, \sigma_{R}^{2}, \sigma_{H}^{2}, \sigma_{D}^{2}, \sigma_{V}^{2},\sigma_{WE}^{2},\sigma_{WC}^{2},\sigma_{WS}^{2}, \sigma_{W}^{2}, \sigma_{Z}^{2}, \sigma_{E}^{2}, \sigma_{F}^{2}, \sigma_{C}^{2},\\ \sigma_{O}^{2}, \sigma_{S}^{2}, \sigma_{U}^{2}) 
\propto\\
p(R| W, Z, \sigma_{R}^{2}) %
p(H| E, F, \sigma_{H}^{2}) %
p(D| C, O, \sigma_{D}^{2}) %
p(V| S, U, \sigma_{V}^{2}) 
p(W| E, \sigma_{WE}^{2}) % 
p(W| C, \sigma_{WC}^{2}) \\
p(W| S, \sigma_{WS}^{2}) %
p(W| \sigma_{W}^{2}) %
p(Z| \sigma_{Z}^{2}) %
p(E| \sigma_{E}^{2}) %
p(F| \sigma_{F}^{2}) %
p(C| \sigma_{C}^{2}) %
p(O| \sigma_{O}^{2}) %
p(S| \sigma_{S}^{2}) %
p(U| \sigma_{U}^{2}). %
\end{split}
\label{eq:FUSION1}
\end{equation}

To calculate the maximum posterior estimation, we get the log of above posterior probability distribution~\cite{jamali2010matrix}.
Derivation of log-posterior probability of Eq.~\ref{eq:FUSION1} is omitted due to space limitation.
Maximizing the  log-posterior probability  with the hyper-parameters (i.e., the observation noise variance and prior variances) is equivalent to minimizing the  objective function Eq.~\ref{eq:FUSION3},
where, 
$\|$.$\|_{F}$ denotes the Frobenius norm.  
$I_{ij}^{*}$ is the indicator function that is equal to 1 if user $U_{i}$ has information regarding item $P_{j}$, otherwise $I_{ij}^{*}$ = 0. 
As for example, $I_{ij}^{H}$ is  equal to 1 if user $U_{i}$ gains helpfulness score on item $P_{j}$, otherwise $I_{ij}^{H}$ = 0. 
$n_{e_{i}}$ is the total number of ratings for which user $U_{i}$ gains helpfulness score  and $n_{f_{j}}$ is the total number of users who gain  helpfulness score  for the rating on item $P_{j}$.
$n_{c_{i}}$ is the total number of ratings for which  $U_{i}$ gains centrality score  and $n_{o_{j}}$ is the total number of users who gain  centrality score  for the rating on $P_{j}$.
Here $n_{s_{i}}$ is the total number of items, those are viewed by user $U_{i}$ and $n_{u_{j}}$ is the total number of users who view the item $P_{j}$
and
\begin{equation*}
% \small
\begin{split}
\lambda_{H} = \dfrac{\sigma_{R}^{2}}{\sigma_{H}^{2}}
\ ; \ \lambda_{D} = \dfrac{\sigma_{R}^{2}}{\sigma_{D}^{2}}
\ ;\  \lambda_{V} = \dfrac{\sigma_{R}^{2}}{\sigma_{V}^{2}}
\ ; \ \lambda_{WE} = \dfrac{\sigma_{R}^{2}}{\sigma_{WE}^{2}}
\ ; \ \lambda_{WC} = \dfrac{\sigma_{R}^{2}}{\sigma_{WC}^{2}}
\ ; \ \lambda_{WS} = \dfrac{\sigma_{R}^{2}}{\sigma_{WS}^{2}}
\ ; \ \lambda_{W} = \dfrac{\sigma_{R}^{2}}{\sigma_{W}^{2}}\\
\ ; \ \lambda_{Z} = \dfrac{\sigma_{R}^{2}}{\sigma_{Z}^{2}}
\ ; \ \lambda_{E} = \dfrac{\sigma_{R}^{2}}{\sigma_{E}^{2}}
\ ; \ \lambda_{F} = \dfrac{\sigma_{R}^{2}}{\sigma_{F}^{2}}
\ ; \ \lambda_{C} = \dfrac{\sigma_{R}^{2}}{\sigma_{C}^{2}}
\ ; \ \lambda_{O} = \dfrac{\sigma_{R}^{2}}{\sigma_{O}^{2}}
\ ; \ \lambda_{S} = \dfrac{\sigma_{R}^{2}}{\sigma_{S}^{2}}
\ ; \ \lambda_{U} = \dfrac{\sigma_{R}^{2}}{\sigma_{U}^{2}}.
\end{split}
\label{eq:parameter}
\end{equation*}

% \section{Generalized Model }

% \begin{equation}
% \begin{split}
% \Phi = \min\limits_{W, Z}   \dfrac{1}{2}\sum_{i=1}^{n}\sum_{j=1}^{m}
% I_{ij}^{R}(R_{ij}-g(W_{i}^T Z_{j}))^2
%  + \sum_{i=1}^{n} \sum_{j=1}^{m}  \bigg[ \sum_{k=1}^{L}\dfrac{\lambda_{{H}^{(k)}}}{•2}I_{ij}^{{H}^{(k)}}        
% (H_{ij}^{(k)}-g(^{(k)}E_{i}^{T}F_{j}^{(k)}))^{2}\bigg]\\
% +\sum_{i=1}^{n}   \bigg[ \sum_{k=1}^{L}\dfrac{\lambda_{W{^{(k)}E}}}{2}\parallel W_{i}-^{(k)}E_{i} \parallel_{F}^{2}\bigg]
% +\sum_{i=1}^{n} \dfrac{\lambda_{W}}{2}n_{w_{i}}\parallel W_{i} \parallel_{F}^{2} 
% +\sum_{j=1}^{m} \dfrac{\lambda_{Z}}{2}n_{z_{j}}\parallel Z_{j} \parallel_{F}^{2} \\
% +\sum_{i=1}^{n}   \bigg[ \sum_{k=1}^{L}\dfrac{\lambda_{^{(k)}E}}{2}n_{^{(k)}e_{i}} \parallel ^{(k)}E_{i}  \parallel_{F}^{2} \bigg]
% +\sum_{j=1}^{m}   \bigg[ \sum_{k=1}^{L}\dfrac{\lambda_{F^{(k)}}}{2}n_{f_{j}^{(k)}} \parallel F_{j}^{(k)}  \parallel_{F}^{2} \bigg]
% \end{split}
% \label{eq:G}
% \end{equation}
%
%
%
% where,
% \begin{equation*}
%  \begin{split}
%  \lambda_{H^{(k)}} = \dfrac{\sigma_{R}^{2}}{\sigma_{H^{(k)}}^{2}} ;
% %
% \lambda_{W^{(k)}E} = \dfrac{\sigma_{R}^{2}}{\sigma_{W^{(k)}E}^{2}} ;
% %
% \lambda_{W} = \dfrac{\sigma_{R}^{2}}{\sigma_{W}^{2}} ;
% %
% \lambda_{Z} = \dfrac{\sigma_{R}^{2}}{\sigma_{Z}^{2}} ;
% %
% \lambda_{^{(k)}E} = \dfrac{\sigma_{R}^{2}}{\sigma_{^{(k)}E}^{2}} ;
% %
% \lambda_{F^{(k)}} = \dfrac{\sigma_{R}^{2}}{\sigma_{F^{(k)}}^{2}}  
% %
% \end{split}
% \end{equation*}
% and $H_{ij}^{(k)}$ is $k^{th}$ parameter (explicit or implicit) and factorize into $^{(k)}E_{i}$ and  $F_{j}^{(k)}$ 
%
The objective function as follows:
\begin{equation}
% \small
\begin{split}
\Phi = \min\limits_{W, Z}   \dfrac{1}{2}\sum_{i=1}^{n}\sum_{j=1}^{m}
I_{ij}^{R}(R_{ij}-g(W_{i}^T Z_{j}))^2
+ \dfrac{\lambda_{H}}{2}\sum_{i=1}^{n}\sum_{j=1}^{m}
I_{ij}^{H}(H_{ij}-g(E_{i}^T F_{j}))^2\\
+ \dfrac{\lambda_{D}}{2}\sum_{i=1}^{n}\sum_{j=1}^{m}
I_{ij}^{D}(D_{ij}-g(C_{i}^T O_{j}))^2
+ \dfrac{\lambda_{V}}{2}\sum_{i=1}^{n}\sum_{j=1}^{m}
I_{ij}^{V}(V_{ij}-g(S_{i}^T U_{j}))^2
+ \dfrac{\lambda_{WE}}{2}\sum_{i=1}^{n} \parallel W_{i} - E_{i} \parallel_{F}^{2}\\
+ \dfrac{\lambda_{WC}}{2}\sum_{i=1}^{n} \parallel W_{i} - C_{i} \parallel_{F}^{2}
+ \dfrac{\lambda_{WS}}{2}\sum_{i=1}^{n} \parallel W_{i} - S_{i} \parallel_{F}^{2}
+\dfrac{\lambda_{W}}{2}\sum_{i=1}^{n} n_{w_{i}}\parallel W_{i} \parallel_{F}^{2}
+\dfrac{\lambda_{Z}}{2}\sum_{j=1}^{m} n_{z_{j}}\parallel Z_{j} \parallel_{F}^{2}\\
+
\dfrac{\lambda_{E}}{2}\sum_{i=1}^{n} n_{e_{i}}\parallel E_{i} \parallel_{F}^{2}
+
\dfrac{\lambda_{F}}{2}\sum_{j=1}^{m} n_{f_{j}}\parallel F_{j} \parallel_{F}^{2}
+
\dfrac{\lambda_{C}}{2}\sum_{i=1}^{n} n_{c_{i}}\parallel C_{i} \parallel_{F}^{2}
+\dfrac{\lambda_{O}}{2}\sum_{j=1}^{m} n_{o_{j}}\parallel O_{j} \parallel_{F}^{2}\\
+\dfrac{\lambda_{S}}{2}\sum_{i=1}^{n} n_{s_{i}}\parallel S_{i} \parallel_{F}^{2}
+\dfrac{\lambda_{U}}{2}\sum_{j=1}^{m} n_{u_{j}}\parallel U_{j} \parallel_{F}^{2}.
\end{split}
\label{eq:FUSION3}
\end{equation} 

\noindent\textbf{Training of $RHCV-PMF$ Model : }
Gradient descent algorithm is used to train our  model, i.e., to minimize the above objective function.
The gradients of $\Phi$ (Eq.~\ref{eq:FUSION3}) with respect to $W_{i}$, $Z_{j}$, $E_{i}$, $F_{j}$, $C_{i}$, $O_{j}$, $S_{i}$ and $U_{j}$  are presented as follows:

\begin{equation}\label{eq:gradient}
% \small
\begin{split}
\frac{\partial \Phi  }{\partial W_{i}}=
\sum_{i=1}^{n}\sum_{j=1}^{m}
I_{ij}^{R} (g'(W_{i}^{T}Z_{j})(g(W_{i}^{T}Z_{j}) - R_{ij})Z_{j})
+\sum_{i=1}^{n}   \bigg[ {\lambda_{WE}} \bigg( W_{i}-E_{i} \bigg)
+ {\lambda_{WC}} \bigg( W_{i}-C_{i} \bigg)\\
+ {\lambda_{WS}} \bigg( W_{i}-S_{i} \bigg)
\bigg] 
 +\sum_{i=1}^{n} {\lambda_{W}}n_{w_{i}} W_{i},
\end{split}
\end{equation}
where, $g'(.)$ is the derivative of tanh function and  
$W_{i}$ is updated as 
\begin{equation}\label{eq:upW}
% \small
%\begin{split}
W_{i}=W_{i}-\varepsilon\frac{\partial \Phi }{\partial W_{i}},
%\end{split}
\end{equation}
where $\varepsilon$ is  learning rate.
Similarly, we evaluate $\frac{\partial \Phi  }{\partial Z_{j}}$,
$\frac{\partial \Phi  }{\partial E_{i}}$,
$\frac{\partial \Phi  }{\partial F_{j}}$,
$\frac{\partial \Phi  }{\partial C_{i}}$,
$\frac{\partial \Phi  }{\partial O_{j}}$,
$\frac{\partial \Phi  }{\partial S_{i}}$,
$\frac{\partial \Phi  }{\partial U_{j}}$, 
update the parameters and 
respective equations are omitted due to space limitation.

\noindent\textbf{Rating Prediction :}
While $\Phi$ has not converged, compute the gradients and update the parameters.
Finally we evaluate predicted rating $\hat R_{ij} = g(W_{i}^{T}Z_{j})$. 
The range of  $\hat R_{ij}$ is -1 to +1 scale and it is mapped into +1 to +5 scale using Eq.~\ref{scale}.
\section{Experiments}
\textbf{Data Statistics:} Our  models are applied on  Amazon.com online review dataset collected by~\cite{amazon} with different datasets on electronics, books, music etc. %and MovieLens-1M ~\cite{harper2016movielens}.
The statistics of the dataset are shown in Table \ref{tab:datast}. 
% \noindent \textbf{Amazon.com:} This data set provides all the required information about each review to build a corresponding review network. 
% %
% This dataset contains:
% $``reviewerID"$, $``asin"$, $``reviewerName"$, $``helpful"$, $``reviewText"$,$ ``overall"$, $``summary"$,\\
% $ ``unixReviewTime"$ ,$``reviewTime"$.
% %
% Here $``asin"$ means item id and  ``$overall"$ denotes rating value.
%
% Here is a sample review entry in the data set: \\
%
%Here $``asin"$ means product id and  ``$overall"$ denotes rating value.
%
% Helpfulness score is taken form ``$helpful"$ attribute and ``helpful": [2, 3] means two customer think that the review is helpful and one customer feels it is not helpful.
% %
% We identify the brand name from $``reviewText"$ field of the review.
%
%\noindent \textbf{MovieLens-1M:} This movie ratings dataset has been widely used to evaluate recommendation system.
%%
%This dataset contains $``UserID"$, $``MovieID"$, $``Rating"$ and $"Timestamp"$ where each user has at least 20 ratings.
%%
%This dataset does not contain any $``helpful"$ parameter.
%
%So, for this dataset $SPS$ is evaluated based on users' ratings and centrality value.
%
%We have taken only rating value to evaluate sentiment score.
%
% We  scale users' ratings.
% %
% We scale ratings from $-2$ to $+2$ i.e. $(-2, -1, 0, +1, +2)$. 
% %
% For example, Amazon.com ratings span from $1$ to $5$.
% %
% So in this case a rating of $5$ would be scaled down to $+2$ and $1$ would be $-2$.
% %
% %
% %
% So we scaled rating from $-2$ to $+2$ (scale of negative to positive).
Amazon.com dataset is extremely sparse. 
The sparseness of the datasets would clearly deteriorate the result  of most exiting recommender systems but our proposed models overcome it.
  
\begin{table} [hbtp]
%\addtolength{\tabcolsep}{.001pt}
\centering
\begin{minipage}{.5\linewidth}
\caption{Statistics of the dataset} 
%\footnotesize
\scalebox{0.63}{
\begin{tabular}{|c c c c c |}
\cline{1-5}
\multirow{1}{*}{\textbf{Dataset}} &\multirow{1}{*}{\textbf{\# users}} & \multirow{1}{*}{\textbf{\# items}}\multirow{1}{*}& {\textbf{\# reviews/ ratings}}\multirow{1}{*}& {\textbf{Sparsity}}   \\ \cline{1-5}
Electronics & 811,034 & 82,067 & 1,241,778&99.998\% \\ %\cline{1-5}

Books & 2,588,991 & 929,264 & 12,886,488&99.999\% \\ %\cline{1-5}

Music & 1,134,684 & 556,814 & 6,396,350 & 99.998\% \\ %\cline{1-5}

Movies and TV & 1,224,267 & 212,836 & 7,850,072 & 99.996\% \\ %\cline{1-5}
Home and  Kitchen & 644509 & 79006 & 991794 & 99.999\% \\ 
% Clothing and Accessories & 128794 & 66370 & 581933 &99.993\% \\ 
Amazon Instant Video & 312930 & 22204 & 717651 & 99.989\% \\ \cline{1-5}
\end{tabular} 
}
\label{tab:datast}
\end{minipage}
\begin{minipage}{0.45\linewidth}
 %  \begin{table} [hbtp]
%\addtolength{\tabcolsep}{.001pt}
\centering
\caption{Comparison MSE results for Amazon.com online reviews dataset on model $RHCV-PMF$ with different settings of dimensionality $K$ (not large number).}
%\footnotesize
\scalebox{0.8}{
\begin{tabular}{|c| c c  c c|}
\cline{1-5}
\multirow{1}{*}{\textbf{Dataset}}
&\multirow{1}{*}{\textbf{$K=5$}}
& \multirow{1}{*}{\textbf{$K=10$}}
&\multirow{1}{*} {\textbf{$K=15$}}  
&\multirow{1}{*} {\textbf{$K=20$}} \\ \cline{1-5}
Electronics & 0.914&0.918 &0.922& 0.925  \\ \cline{1-5}
Books & 0.912& 0.822& 0.811&  0.811 \\ \cline{1-5}
Music & 0.701& 0.701& 0.699&0.698  \\ \cline{1-5}
Movies and TV & 0.792& 0.791& 0.767&0.733  \\ \cline{1-5}
\end{tabular} 
}
\label{tab:K}
\end{minipage}
\end{table}
% \caption{MSE results of various model based on Amazon.com online reviews dataset, where rating scale 1 to 5 \cite{baseline1} and set $K$ = 5 and 80\% as Training Dataset}.
% %\footnotesize
% \scalebox{.7}{
% \begin{tabular}{|c c c c c c |}
% \cline{1-6}
% \multirow{1}{*}{\textbf{Dataset}} &\multirow{1}{*}{\textbf{\ MF}} & \multirow{1}{*}{\textbf{\ LDAMF}}\multirow{1}{*}& {\textbf{\ CTR}} & \multirow{1}{*}{\textbf{\ HFT}} &\multirow{1}{*} {\textbf{\ RMR}}  \\ \cline{1-6}
% Electronics & 1.828 & 1.823 & 1.764 &1.722 &1.722  \\ %\cline{1-6}

% Books & 1.107 & 1.109 & 1.106 &1.138 &1.113  \\ %\cline{1-6}

% Music & 0.956 & 0.958 & 0.959 & 0.980 & 0.959  \\ %\cline{1-6}

% Movies and TV & 1.119 & 1.117 & 1.114 &1.119 &1.120 \\ \cline{1-6}
% \end{tabular} 
% }
% \label{tab:baseline1}
% \end{minipage}
% \end{table}

\noindent\textbf{Evaluation Metric and baseline methods :}
One widely used evaluation metric $i.e.$, mean square error (MSE) is considered to evaluate the performance of  our models.
%
% MAE is defined as
 % \begin{equation}\label{eq:MAE}
% MAE= \dfrac{\sum_{(i,j)\epsilon \tau} \vert \hat {sps_{ij}}- sps_{ij}\mid }{\mid\tau\mid}
% \end{equation}
% MSE is defined as
%  \begin{equation}\label{eq:MSE}
% MSE= \dfrac{\sum_{(i,j)\epsilon \tau} ( \hat {sps_{ij}}- sps_{ij})^{2} }{\mid\tau\mid}
% \end{equation}
% Here, average percentage of error is defined as
% %
% \begin{equation}\label{eq:error}
% Avg \ \% \ of \  error= \dfrac{\sum_{(i,j)\epsilon \tau}\mid\dfrac{ \hat {sps_{ij}}- sps_{ij} }{sps_{ij}}|\times100}{ \mid\tau\mid},
% \end{equation}
% where $\tau$ denotes the set of $SPS$ we want to predict, $sps_{ij}$ denotes the sps that user $u_{j}$ gains for item $p_{i}$ and $\hat {sps_{ij}}$ is the predicted $sps$. 
% \subsection{Baseline Methods}
Five previously proposed models $i.e.$, Matrix Factorization (MF)~\cite{mnih2008probabilistic}, Latent Dirichlet Allocation using MF (LDAMF)~\cite{amazon}, Collaborative Topic Regression (CTR)~\cite{CTR}, Hidden Factors as Topics (HFT)~\cite{amazon}, Ratings Meet Review (RMR)~\cite {baseline1} and Modeling on product image and ``also-viewed'' product information (VMCF)~\cite{park2017also} are considered as baseline models.
All these models  are applied on  Amazon.com dataset and  predict rating of a user for a particular item.
In table \ref{tab:baselinecom}, $2^{nd}$ to $7^{th}$ column  shows $MSE$ results of the baseline models. 
\noindent\textbf{Parameters :}
% For ItemCF, we set the number of neighbors to 20.
%
For our model $RHCV-PMF$, we perform our experiment with $\lambda_{H}, \lambda_{D}$, $\lambda_{V}$ $\in$ \{0.1, 0.3, 0.5\},
$\lambda_{W}, \lambda_{Z}, \lambda_{E}, \lambda_{F}, \lambda_{C}, \lambda_{O}$, $\lambda_{S}, \lambda_{U}$ $\in$ [0.1, 1] and $\lambda_{WE}, \lambda_{WC},\\
\lambda_{WS}$ $\in$ [0.1, 0.5] while $K$ is fixed to 5. 
As a result, while $\lambda_{H} = \lambda_{D}$ = $\lambda_{V}$ = 0.2, 
$\lambda_{W} = \lambda_{Z} =\lambda_{E} =\lambda_{F} =\lambda_{C} =\lambda_{O} = \lambda_{S} =\lambda_{U}$ = 0.1 manifest the best performance for all experimental datasets.
$\lambda_{WE} =\lambda_{WC}=\lambda_{WS}$ = 0.2 yield the best performance for Electronics and Music dataset.
$\lambda_{WE} =\lambda_{WC}=\lambda_{WS}$ = 0.3 yield the best performance for Books, Movies and TV dataset.
Details of observation with different parameters are omitted due to space limitation.

\noindent\textbf{Evaluation :}
We randomly select 80 \% of the user ratings dataset for training, and the remaining 20 \% is used for testing. 
Random sampling is independently conducted five times and we perform our experiment on the baseline models and our models.
%
% The training of the baseline models as shown in Table \ref{tab:baselinecom}, are described in \cite{amazon}
%
In Table \ref{tab:baselinecom}, comparison of MSE results between baseline models and our models based on Amazon.com online review dataset are shown.
We use $K$ = 5 for all models.
In this table, $1^{st}$ column indicates different types of dataset,
$2^{nd}$ to $7^{th}$ column shows the performance of base-line models.
Last three columns show the performance of our models.
For each dataset, our models perform better significantly.
$RHCV-PMF$ model, fusion of $RHC-PMF$ and $RV-PMF$, performs better compare to  $RHC-PMF$ and $RV-PMF$.
%
% It should be noted that the results for Music, Movies and TV dataset are generally better than the results of Other three datasets for all models, because of 
\begin{table} [hbtp]
%\addtolength{\tabcolsep}{.001pt}
\centering
\caption{Comparison MSE results between baseline models (mention in $2^{nd}$ to $7^{th}$ column) and our models (mention $8^{th}$ to $10^{th}$ column) based on Amazon.com online reviews dataset, where rating scale 1 to 5, $K$ = 5  and 80\% as Training Dataset is  used.}
%\footnotesize
\scalebox{0.8}{
\begin{tabular}{|c| c c  c c c c| c| c |c| }
\cline{1-10}
\multirow{1}{*}{\textbf{Dataset}}  &\multirow{1}{*}{\textbf{\ MF}} & \multirow{1}{*}{\textbf{\ LDAMF}}\multirow{1}{*}& {\textbf{\ CTR}} & \multirow{1}{*}{\textbf{\ HFT}} &\multirow{1}{*} {\textbf{\ RMR}} &\multirow{1}{*} {\textbf{\ VMCF}}
&\multirow{1}{*} {\textbf{\ RHC-PMF}}
&\multirow{1}{*} {\textbf{\ RV-PMF}}
&\multirow{1}{*} {\textbf{\ RHCV-PMF}}  \\ \cline{1-10}
Electronics & 1.828 & 1.823 & 1.764 &1.722 &1.722& 1.521&1.233 &1.523 &0.914  \\ %\cline{1-6}

Books & 1.107 & 1.109 & 1.106 &1.138 &1.113& 1.021&0.987&1.034& 0.912 \\ %\cline{1-6}

Music & 0.956 & 0.958 & 0.959 & 0.980 & 0.959&0.950 &0.899 &0.951&0.701 \\ %\cline{1-6}

Movies and TV & 1.119 & 1.117 & 1.114 &1.119 &1.120& 1.028&0.917&1.029& 0.792\\  
Home and Kitchen & 1.628 & 1.610 & 1.577 &1.531 &1.501&1.373 &1.133 &1.377&1.191\\ 
% Clothing and Accessories& 0.393& 0.406 & 0.355 &0.349 &0.336& 0.622\\ 
Amazon Instant Video  &1.330 & 1.328& 1.291 &1.260 &1.270&1.269 &1.145&1.271& 1.102\\ \cline{1-10}
\end{tabular} 
}
\label{tab:baselinecom}
\end{table}

 \noindent\textbf {Different settings of dimensionality $K$:}
Increasing $K$ (not large value) should add more flexibility to a model and as a result, it should improve the performance.
But in Table \ref{tab:K}, we notice that increment of $K$ for Books, Music, Movies and TV improved the result but for Electronics dataset, it did not improve the result.
 For this  type of contradictory results,  our opinion is that Electronics dataset is smaller than the other three datasets and increasing $K$ leads to more parameter in the model which leads to overfitting. 
 \begin{table} [hbtp]
%\addtolength{\tabcolsep}{.001pt}
\centering
\caption{Comparison MSE results of cold-start items and users between baseline models, $ItemCF$~\cite{CF2}, $VMCF$~\cite{park2017also} and our model $RHCV-PMF$ based on Amazon.com online reviews dataset, where rating scale 1 to 5 and $K$ = 5.  }
%\footnotesize
\scalebox{.8}{
\begin{tabular}{|c| c| c | c| c |c |c|}
\cline{1-7}
\multirow{2}{*}{\textbf{Dataset}} &\multicolumn{3}{l|}{\textbf{Cold start items}}
&\multicolumn{3}{l|}{\textbf{Cold start users}}
\\ \cline{2-7}
&\multirow{1}{*}{\textbf{\ ItemCF}}
& \multirow{1}{*}{\textbf{\ VMCF}}
&\multirow{1}{*} {\textbf{\ RHCV-PMF}}
&\multirow{1}{*}{\textbf{\ ItemCF}}
& \multirow{1}{*}{\textbf{\ VMCF}}
&\multirow{1}{*} {\textbf{\ RHCV-PMF}}
\\ \cline{1-7}

Electronics & 1.957&1.547&1.112&1.833&1.534&1.217  \\ \cline{1-7}
Books & 1.277& 1.212& 1.133&1.298 &1.227&1.109  \\ \cline{1-7}
Music & 1.134& 1.116& 0.981&1.155&1.234&1.107  \\ \cline{1-7}
Movies and TV & 1.533& 1.487& 1.177& 1.567&1.503& 1.113\\ \cline{1-7}
\end{tabular} 
}
\label{tab:coldstar}
\end{table}

\noindent\textbf {Performance on cold start users and items :}
We also evaluate the performance of our model $RHCV-PMF$  for cold-start users and cold-start items as shown in Table \ref{tab:coldstar}.
For cold start items we compare our performance with two baseline models $ItemCF$ \cite{CF2}, $VMCF$ \cite{park2017also} that are mentioned in $2^{nd}$ and $3^{rd}$ column,
and for cold start users we compare our model performance with two same baseline models that are mentioned in $5^{th}$ and $6^{th}$ column.
The baseline models and our models are applied on Amazon.com dataset.
We consider the users who have expressed less than four ratings as cold start users and for items, which have received less than four ratings in the training dataset, are considers as cold start items.
Our observation is that 50 \% users and 40 \% items are cold start users and items, respectively.
Our model performs better than baseline models due to the consideration of both implicit and explicit feedback for cold start users and items also.

\section{Conclusion and Future Work}
In our research work, we investigate Probabilistic Matrix Factorization ($PMF$) based model  for recommendation system, that considers explicit feedbacks and implicit feedbacks. 
In our work, it is proved that fusion of explicit feedbacks and implicit feedbacks is really effective.
%
% We also generalize our model $RHCV-PMF$ for $L$ number of parameters.
%
In our investigation, it is clearly proved that our model is really effective for cold start users and items also.

% In future, there are several interesting directions that need further investigation.
%
In future we would like to apply our dataset on other models.
We would like to experiment on other online merchandise companies' datasets.
Several online merchandise companies connect with social medias and users share their reviews in social medias.
Now social networks are available in social medias and allow sources for suitable recommendation.
We would like to investigate if social networks can be utilized to learn users' preference areas and item rating activity.

\bibliographystyle{spmpsci}
%\bibliography{sample-bibliography}

%\begin{thebibliography}{6}
%%
%
%\bibitem {smit:wat}
%Smith, T.F., Waterman, M.S.: Identification of common molecular subsequences.
%J. Mol. Biol. 147, 195?197 (1981). \url{doi:10.1016/0022-2836(81)90087-5}
%
%\bibitem {may:ehr:stein}
%May, P., Ehrlich, H.-C., Steinke, T.: ZIB structure prediction pipeline:
%composing a complex biological workflow through web services.
%In: Nagel, W.E., Walter, W.V., Lehner, W. (eds.) Euro-Par 2006.
%LNCS, vol. 4128, pp. 1148?1158. Springer, Heidelberg (2006).
%\url{doi:10.1007/11823285_121}
%
%\bibitem {fost:kes}
%Foster, I., Kesselman, C.: The Grid: Blueprint for a New Computing Infrastructure.
%Morgan Kaufmann, San Francisco (1999)
%
%\bibitem {czaj:fitz}
%Czajkowski, K., Fitzgerald, S., Foster, I., Kesselman, C.: Grid information services
%for distributed resource sharing. In: 10th IEEE International Symposium
%on High Performance Distributed Computing, pp. 181?184. IEEE Press, New York (2001).
%\url{doi: 10.1109/HPDC.2001.945188}
%
%\bibitem {fo:kes:nic:tue}
%Foster, I., Kesselman, C., Nick, J., Tuecke, S.: The physiology of the grid: an open grid services architecture for distributed systems integration. Technical report, Global Grid
%Forum (2002)
%
%\bibitem {onlyurl}
%National Center for Biotechnology Information. \url{http://www.ncbi.nlm.nih.gov}
%
%
%\end{thebibliography}
\end{document}